# Pressure-induced Superconductivity in AgSbTe$_2$

Sudaice Kazibwe[1], Bishnu Karki[1], Wencheng Lu[2], Zhongxin Liang[1], Minghong Sui[1], Melissa Gooch[1], Zhifeng Ren[1], Pavan Hosur[1], Timothy A. Strobel[2*], Ching-Wu Chu[1*], Liangzi Deng[1*]

1. Department of Physics and Texas Center for Superconductivity, University of Houston, Houston, Texas 77204, USA

2. Earth and Planets Laboratory, Carnegie Institution for Science, Washington, DC 20015, USA

* tstrobel@carnegiescience.edu, cwchu2@central.uh.edu, ldeng2@central.uh.edu

**Abstract**

AgSbTe$_2$ is a well-known thermoelectric material with a high Seebeck coefficient and intrinsically low thermal conductivity, but its behavior under pressure remains largely unexplored. Here we report a systematic investigation of the structural, electronic, and transport properties of non-stoichiometric AgSbTe$_2$ under high pressure. At ambient pressure, the material can be described as having a cubic crystal structure that remains stable up to 21.7 GPa beyond which it loses long-range structural order, while its crystal system fully recovers upon decompression. Remarkably, superconductivity emerges at a very low pressure of 0.38 GPa with an onset superconducting critical temperature ($T_c$) of 3.2 K. $T_c$ increases with increasing pressure, reaching 6.9 K at 31.9 GPa, and peaks at 7.4 K during decompression. Magnetic-field-dependent transport measurements and electronic structure calculations reveal an evolution of the superconducting state driven by an enhanced electronic density of states at the Fermi level under compression. Our findings uncover pressure-induced superconductivity in AgSbTe$_2$ and demonstrate that pressure can effectively tune the electronic ground state of thermoelectric materials, extending their functionality beyond thermoelectric energy conversion.

## 1. Introduction

AgSbTe$_2$ is a narrow band gap I-V-VI$_2$ semiconductor that has attracted sustained interest due to its impressive thermoelectric performance in the mid-temperature range [1-3]. Its high Seebeck coefficient combined with an intrinsically low lattice thermal conductivity (0.5-0.7 W/mK) [4-7] results in a remarkable thermoelectric figure of merit (ZT) between 300 and 600 K [3], with a maximum ZT of 2.6 at 573 K achieved *via* partial substitution of Cd on the Sb cation sublattice [8]. The notable low lattice thermal conductivity has been attributed to the spontaneous formation of nanostructures and anharmonic bonding in AgSbTe$_2$ [4,9].

Despite extensive studies of its thermoelectric properties, the crystal structure of AgSbTe$_2$ remains debated. In an early study by Geller and Wenick, they proposed a cubic rock-salt structure (space group $Fm\bar{3}m$), with statistical disorder of Ag and Sb on the 4a Wyckoff site and Te occupying the 4b site [10]. More recently, B. Sun *et al.* suggested a rhombohedral structure ($R\bar{3}m$) at ambient pressure (AP) [11], while Quarez *et al.* reported several possible ordered superstructures including *P4/mmm*, $Pm\bar{3}m$, and $R\bar{3}m$ symmetries [12], with most structures probed by powder X-ray diffraction, except in the study by B. Sun *et al.* First-principles calculations further predict ordered configurations such as the $L1_1$ and D4 cation superstructures [13-15]. These contrasting results indicate that powder diffraction cannot unambiguously determine the exact crystal structure. However, Raman and NMR measurements, together with electric field gradient calculations, strongly support a rhombohedral phase at AP [11].

Under pressure, $AgSbTe_2$ has been reported to undergo a B1 (NaCl-type) to B2 (CsCl-type) structural transition between 17 and 26 GPa *via* an intermediate amorphous phase, which can be quenched to ambient conditions [16]. However, Ko *et al.* observed pressure-induced amorphization near 24.6 GPa followed by recrystallization above 49.2 GPa, without clear evidence for the formation of the B2 phase [17]. These discrepancies suggest that the crystal structural evolution of $AgSbTe_2$ under pressure remains controversial and warrants systematic reinvestigation, particularly in the off-stoichiometric composition $Ag_{0.9}Sb_{1.1}Te_{2.1}$, in which decomposition of stoichiometric $AgSbTe_2$ into secondary phases ($Ag_2Te$ and $Sb_2Te_3$) below 633 K can be effectively suppressed.

In parallel, pressure-induced superconductivity has been widely reported in ternary chalcogenides. Many layered $AB_2X_4$ compounds (A =Ge, Sn, and Pb; B =Sb and Bi; X =Se and Te) are topological insulators at AP and exhibit superconductivity under compression. Notably, pressure-induced topological superconductivity has been observed in $PbBi_2Te_4$ [18], $GeSb_2Te_4$ [19], $SnSb_2Te_4$ [20], and $SnBi_2Se_4$ [21]. Superconductivity under pressure has also been reported in non-layered chalcogenides such as $Ta_2PdS_6$ [22] and $Ta_2PdSe_6$ [23]. Since superconductivity in compressed narrow-gap and semi metallic systems is often associated with pressure-induced modifications of the electronic band structure [24,25], it is natural to explore whether similar phenomena may emerge in $AgSbTe_2$-based systems, given their narrow band gap and structural complexity.

Here, we present a comprehensive high-pressure study of $Ag_{0.9}Sb_{1.1}Te_{2.1}$ (AST) combining synchrotron X-ray diffraction (XRD) and electrical transport measurements up to 55 GPa. At 0.5 GPa, AST can be described by the cubic structure that remains stable up to 21.7 GPa. Under further increasing pressure, the diffraction peaks progressively broaden and weaken, indicating the loss of long-range structural order that persists to 25.4 GPa, where only the (200) reflection remains discernible. Upon decompression, the original crystalline phase fully recovers, demonstrating reversible pressure-induced amorphization-like behavior and revealing the intrinsic structural instability of the system.

Remarkably, superconductivity emerges at 0.38 GPa with an onset superconducting critical temperature ($T_c$) of 3.2 K. A zero-resistance state is achieved at 12.3 GPa, confirming the discovery of a superconducting phase in AST. With further compression, $T_c$ increases and reaches a maximum value of 6.9 K at 31.9 GPa. $T_c$ is further enhanced during decompression, peaking at 7.4 K at 26.7 GPa. To investigate the microscopic origin of this superconducting phase, electronic band-structure calculations were performed for the compound at AP and under applied pressure. The calculations reveal an increase in the electronic density of states from ambient pressure to 25 GPa, along with the emergence of imaginary phonon modes in the 25-30 GPa range, consistent with the experimentally observed transition in which the material loses its long-range crystalline order.

The observation of superconductivity in AST provides new insights into the mechanisms of superconductivity in the chalcogenides and extends the functional scope of $AgSbTe_2$-based materials beyond thermoelectricity, opening new opportunities for exploring emergent quantum phenomena in I-V-VI$_2$ compounds.

## 2. Methodology

2.1 Sample synthesis

Polycrystalline $Ag_{0.9}Sb_{1.1}Te_{2.1}$ was prepared *via* a melt–ball-milling–hot-pressing route. High-purity Ag (99.999%), Sb (99.99%), and Te (99.999%) were weighed according to the nominal composition and sealed in an evacuated quartz tube. The tube was heated to 1173 K over 10 h and maintained at this temperature for an additional 6 h. The melt was then cooled to 723 K at a controlled rate over 8 h and held at this temperature for 48 h to enhance compositional homogeneity, after which rapid water quenching was applied to obtain the ingot. The ingot was subsequently ball-milled into a fine powder and then consolidated into dense pellets using a graphite die (12.7 mm in diameter) by hot pressing at 673 K for 5 min under a constant pressure of 60 MPa. The sample morphology and composition were characterized by scanning electron microscopy (SEM) and energy-dispersive X-ray spectroscopy EDS respectively as shown in Fig. S1c and S1d. Following synthesis, a composition of $Ag_{0.9}Sb_{1.11}Te_{2.08}$, which is close to the initial composition of the mixture, was found.

2.2 XRD under pressure

A rhenium gasket was pre-indented to a thickness of 40 $\mu$m and loaded into a diamond anvil cell (DAC) with a 400 $\mu$m culet diameter. After pre-indentation, a sample chamber was created by drilling a hole through the gasket with a diameter of 180 $\mu$m, after which a flake of the compressed starting precursors with a thickness of ~10 $\mu$m was loaded. After loading the precursor, high-density fluid neon was used as a pressure-transmission medium (PTM) to produce quasi-hydrostatic pressure. Up to ~5 GPa, the pressure was determined using an internal gold diffraction standard prior to the solidification of neon, while above 5 GPa, neon diffraction was employed [26,27]. Focused monochromatic synchrotron XRD measurements were carried out at beamline 13-ID-D (GSECARS) of the Advanced Photon Source, Argonne National Laboratory (run 6, $\lambda$ = 0.3344 Å) using an Eiger2 CdTe 9M detector. The sample–detector distance and other geometric parameters were calibrated using a CeO$_2$ standard, and two-dimensional diffraction patterns were integrated using the DIOPTAS package [28].

2.3 Electrical transport measurements under pressure

Resistivity measurements up to 55 GPa were conducted using a BeCu DAC that is adapted for use in the Quantum Design Physical Property Measurement System (PPMS). Diamonds with a culet size of 300 $\mu$m were used, and the rhenium gasket was insulated with Stycast 2850FT, with cubic boron nitride used as the PTM. After the gasket was preindented, a 130 $\mu$m diameter sample chamber was made in the center of the culet. Samples with a diameter of about 120 $\mu$m were placed in the sample space. The pressure below 10 GPa was determined using the ruby fluorescence scale [29], and above 10 GPa, the diamond Raman scale at room temperature [30] was used. The samples' contacts were arranged in a van der Pauw configuration [31], and data were collected using a Quantum Design PPMS at temperatures down to 2.0 K and under magnetic fields up to 7 T.

2.4 DFT calculations

Electronic structure calculations were performed using density functional theory (DFT) implemented in the Vienna Ab Initio Simulation Package (VASP) [32,33] and the Full Potential Local Orbital (FPLO) code [34]. Crystal structure optimization, hydrostatic pressure application, and phonon calculations were conducted using VASP. Electronic band structure and Fermi surface calculations were performed using FPLO.

For crystal structure optimization, the Perdew-Burke-Ernzerhof generalized-gradient approximation (GGA-PBE) was used to compute the exchange-correlation functional [35]. The reciprocal space was sampled with a $\Gamma$-centered $k$-mesh of size 6 × 6 × 6, along with a plane-wave kinetic energy cutoff of 500 eV. Energy and force convergence criteria were set to $10^{-5}$ eV and $10^{-3}$ eV/Å, respectively. Hydrostatic pressure was applied up to 30 GPa. The crystal structures were fully optimized for both the inner coordinates and the cell parameters at each pressure. All DFT calculations were carried out using a 64-atom unit cell of $AgSbTe_2$. Phonon calculations were performed using the finite-difference method with a 2 × 2 × 2 supercell. The phonopy package [36] was used to plot the phonon dispersions. The electronic structure calculations using FPLO employed the same DFT convergence parameters mentioned earlier, with a k-mesh size of 12 × 12 × 12 for band structure calculations and that of 30 × 30 × 30 for Fermi surface calculations.

## 3. Results and Discussion

### 3.1 Pressure-induced superconductivity

$Ag_{0.9}Sb_{1.1}Te_{2.1}$ shows no signature of superconductivity down to 2 K at AP, as confirmed by temperature-dependent electrical transport and magnetization measurements shown in Fig. S1a and S1b, respectively. The resistivity measurements at AP indicate that the sample shows activated transport behavior, confirming the presence of a narrow band gap in the material. Upon applying pressure up to 0.38 GPa as shown in Fig. 1a, the sample exhibits semiconducting behavior in the normal state, followed by a drop in resistance (R) with an onset $T_c$ of 3.2 K. With increasing pressure, up to 9.8 GPa, the $T_c$ continues to increase, and the R drop becomes more pronounced, accompanied by a complete crossover of the normal-state behavior from semiconducting to metallic. As shown in Fig. 1b, zero R was observed at 12.3 GPa, confirming for the first time the observation of a pressure-induced superconducting phase in this material. R measurements were extended up to 52.6 GPa, yielding a maximum $T_c$ of 6.6 K at 28.7 GPa. Notably, the zero-R temperature increases from 3.2 K at 12.3 GPa to 5.4 K at 40.8 GPa, further demonstrating a pressure-enhanced superconducting phase in AST. The superconducting behavior was reproducible, as shown by the results from additional Run-2 and Run-3 experiments in Fig. S2a, S2b, S2c, and S2d. With increasing pressure, an overall maximum $T_c$ of 6.9 K was obtained at 31.9 GPa during Run-3 (Fig. S2d). Under decompression, as shown in Fig. S2f, a further enhancement of $T_c$ was observed, reaching a maximum of 7.4 K at 26.7 GPa. A direct comparison of the R-T curves obtained during compression and decompression at nearly identical pressures is presented in Fig. S3.

To further characterize the superconducting state, we performed magnetic field–dependent measurements at 40.8 GPa during Run-1. As shown in Fig. 1c, the superconducting transition systematically shifts to lower temperatures with increasing magnetic fields up to 1 T. This field-dependent suppression of $T_c$ confirms that the observed resistance drop originates from superconductivity rather than being a pressure-induced anomaly. Additional field-dependent measurements show that the superconducting state is completely suppressed in a magnetic field of less than 4 T (Fig. S2e). The upper critical field ($\mu_0H_{c2}$) at absolute zero temperature was determined using the empirical Ginzburg–Landau fitting [37], $\mu_0H_{c2}(T) = \mu_0H_{c2}(0)[1 - (t)^2]/[1 + (t)^2]$ for $t = T/T_c$, as shown in Fig. 1d. The extracted upper critical field at different criteria were used to compute the coherence length from $\xi^2(0) = \Phi_0/2\pi\mu_0H_{c2}(0)$, where $\Phi_0 = 2.07 \times 10^{-15}$ Wb is the magnetic flux quantum [38], and the obtained results are summarized in Table 1.

| Criterion | 90% of $R_N$ | 50% of $R_N$ | 10% of $R_N$ |
|---|---|---|---|
| $\mu_0 H_{C2}$(T) | 3.02(3) | 2.66(5) | 2.21(2) |
| $\xi$ (0) (nm) | 10.44 | 11.13 | 12.21 |

Table 1| Summary of the extracted upper critical field and computed coherence length using the Ginzburg-Landau fitting at 90%, 50%, and 10% drops from the normal state resistance.

The obtained upper critical field $H_{c2}$ of 3.02 T is well below the Pauli paramagnetic limiting field, $\mu_oH_{Pauli}$ = 1.86 $T_c$. For $T_c$ = 6.6 K, this yields $\mu_oH_{Pauli}$ = 12.3 T, which is larger than the experimentally measured $H_{c2}$. This suggests that the superconductivity in this compound is primarily suppressed by orbital pair-breaking rather than by spin polarization. The condition $H_{c2}<\mu_oH_{Pauli}$ is consistent with a superconducting state that is not limited by Pauli paramagnetism and may indicate conventional phonon-mediated pairing in AST [39]. Similar behavior has been reported in several chalcogenide superconductors, such as $Sn_{1-x}In_xTe$ [40], which likewise do not approach their Pauli paramagnetic limits.

3.2 XRD under pressure

Fig. 2a shows the pressure dependence of the powder XRD pattern of $Ag_{0.9}Sb_{1.1}Te_{2.1}$, in which a systematic shift of all peaks toward higher 2θ values with increasing pressure, an indication of lattice compression, is observed. We note that the diffraction peak emerging at 4.9 GPa and denoted by a gray star, observed at a 2θ angle of 10.25° and systematically shifting to higher angles with increasing pressure, is due to the solidification of neon, the PTM. Upon further compression up to 21.7 GPa, the diffraction peak intensities gradually decrease. At pressures above 23.6 GPa, the diffraction peaks become asymmetric and their intensities decrease markedly, signaling the loss of long-range crystalline order, and higher angle peaks disappear. This behavior is consistent with previous reports on stoichiometric $AgSbTe_2$ compounds [16,17]. The onset of an amorphous-like phase, characterized by the loss of long-range crystalline order, is consistent with our phonon dispersion calculations, which reveal dynamic instability of the structure in the pressure range of 25-30 GPa. Similar pressure-induced amorphization has been well documented in materials such as α-quartz and coesite, which undergo this transition in the pressure range of 25-35 GPa [41]. Fig. 2b shows the Rietveld refinement at 0.5 GPa. Although, given the similarity between competing structural models, it is difficult to unambiguously determine the exact crystal structure of this compound from powder XRD alone, as recently reported by Sun *et al.* [11] and also evident in our data (Fig. S4a), a close inspection of the low-intensity region of our data reveals that, at 0.5 GPa, the diffraction peaks can be fitted with $Pm\bar{3}m$, $Fd\bar{3}m$ (Fig. S4b), and $R\bar{3}m$ (Fig. S4c). The refinement with $Pm\bar{3}m$ yields a lattice parameter of 6.0627 Å. No diffraction peaks corresponding to $Sb_2Te_3$ or $Ag_2Te$ were detected within the resolution limit of the synchrotron measurements. The absence of characteristic reflections of these phases suggests that any secondary phase content, if present, is below the detection threshold.

Fig. 2c illustrates the evolution of the (200) Bragg diffraction peak during compression from 0.5 to 25.4 GPa. With increasing pressure, the peak progressively shifts toward higher 2θ values, indicating a continuous decrease in interplanar spacing. At 24.5 GPa, a weak and asymmetric peak emerges, while higher-angle peaks disappear, signaling the onset of the loss of long-range crystalline order, which is characteristic of an amorphous phase. The amorphous phase persists up to the maximum applied pressure of 25.4 GPa. The corresponding 2D diffraction images collected at 0.5 GPa and 25.4 GPa (Fig. S5a and S5b, respectively) clearly demonstrate the loss of long-range crystalline order at 25.4 GPa.

Upon decompression back to ambient pressure (AP), the sample recrystallizes into the cubic structure, as shown in Fig. 2d. This reversible behavior is consistent with that previously reported for stoichiometric $AgSbTe_2$ [16]. Fig. S4d also shows a systematic decrease in the lattice parameter and the unit-cell volume with increasing pressure, providing clear evidence of lattice compression.

The absence of any pressure-induced structural phase transition in $Ag_{0.9}Sb_{1.1}Te_{2.1}$ at lower pressures suggests that the observed variation in the superconducting transition temperature, an initial increase followed by a decrease and a subsequent increase over the pressure range from AP to 7 GPa, is likely driven by changes in the electronic structure and/or other, as yet unidentified, mechanisms rather than by a crystallographic transformation.

3.2 $Ag_{0.9}Sb_{1.1}Te_{2.1}$ $T_c$ – P phase diagram

Fig. 3 shows the pressure dependence of the $T_c$ for $Ag_{0.9}Sb_{1.1}Te_{2.1}$ measured over multiple experimental runs. The sample exhibits a non-monotonic variation of $T_c$ with applied pressure up to 8 GPa. With further compression up to 30 GPa, $T_c$ increases at a rate of 0.15 K/GPa (between 8 to 30 GPa). Such $T_c$ enhancement can be qualitatively correlated with an increase in the electronic density of states at the Fermi level, as indicated by the band-structure calculations shown in Fig. 4b and discussed below. With further compression between 30 and 53 GPa, the $T_c$ decreases monotonically at a rate of -0.06 K/GPa. This suppression is likely associated with the loss of the long-range crystalline order, as revealed by the XRD data in Fig. 2a, which may be unfavorable for superconductivity, as well as with the structural instability indicated by the phonon dispersion calculations discussed below. Although small quantitative variations in $T_c$ are observed among different runs, the overall dome-like pressure dependence is reproducible, indicating that the superconducting behavior is indeed intrinsic to the material. The existence of superconductivity within the amorphous phase suggests that it may originate from the residual crystalline domains embedded within a partially amorphous matrix.

Measurements performed during decompression clearly show an enhanced $T_c$, likely due to pressure-induced phonon softening [42]. Upon releasing pressure below 25 GPa, the phonon modes harden, leading to a reduction in $T_c$, and superconductivity vanishes once the system is fully returned to AP (Fig. S2f). A similar decompression-induced enhancement of superconductivity has been reported in related systems [42-44]. A direct comparison of the R-T curves obtained during compression and decompression at nearly identical pressures is presented in Fig. S3.

3.4 DFT calculations

To gain deeper insight into the observed increase in $T_c$ and to investigate the evolution of the electronic band structure under pressure, we performed density functional theory (DFT) calculations from AP up to 30 GPa. Fig. 4a and 4c show the calculated band structure of $AgSbTe_2$, revealing a semimetallic character in which the valence band maximum significantly overlaps with the conduction band minimum by approximately 0.33 eV. This value is close to the 0.3 eV reported by Ye *et al*. [45] and the 0.28 eV reported by Min *et al*. [46] at AP. Overall, our results are in good agreement with several previously reported studies on this compound [13,14,46-50]. Although earlier experimental studies described pristine $AgSbTe_2$ as a semiconductor with a band gap of 0.3-0.38 eV, as determined from optical reflectance measurements [51-53], our PBE-based DFT calculations do not accurately reproduce this band gap. This discrepancy between theory and experiment is likely due to the well-known underestimation of band gaps within the generalized gradient approximation (GGA) for the exchange-correlation potential [54]. Nevertheless, our calculations provide clear and definitive insight into the pressure-induced evolution of the electronic band structure.

With increasing pressure, both the electron and hole pockets expand noticeably, indicating an increase in the density of states at the Fermi level. The states near the Fermi level are predominantly derived from Sb-$5p$ and Te-$5p$ orbitals, as shown in Fig. S6. Fig. 4b shows the calculated pressure-dependent density of states up to 25 GPa. As pressure increases, the enlargement of the electron and hole pockets leads to a progressively more crowded Fermi surface, reflecting the growing number of states contributing at the Fermi level (Fig. S8). The individual bands contributing to the Fermi surface are illustrated in Fig. S7. Overall, the clear enhancement of the electronic density of states with applied pressure correlates well with the observed increase in $T_c$ up to 25 GPa.

Fig. 5 shows the evolution of the phonon dispersion under pressure up to 30 GPa. At pressures from ambient conditions to 25 GPa (Fig. 5a and 5b), the phonon spectra exhibit only positive frequencies, indicating structural stability. Upon further increasing the pressure, imaginary phonon modes emerge between 25 and 30 GPa, as shown in Fig. 5c, signaling a lattice instability. This pressure range coincides with the transition to an amorphous phase observed in our XRD measurements. While phonon frequencies generally harden under compression, the concurrent increase in electronic density of states may enhance the overall electron-phonon coupling strength, potentially contributing to the observed $T_c$ enhancement.

## 4. Conclusions

In summary, we have experimentally observed superconductivity in the non-stoichiometric compound AgSbTe$_2$ under pressure. Superconductivity emerges with a $T_c$ of 3.2 K at 0.38 GPa, with increasing pressure reaches a maximum $T_c$ of 6.9 K at 31.9 GPa, and is enhanced during decompression with a peak $T_c$ of 7.4 K at 26.7 GPa. X-ray diffraction measurements indicate that the sample can be better fitted with the cubic $Pm\bar{3}m$ structure, although other models such as $R\bar{3}m$, $P4/mmm$, and $Fd\bar{3}m$ are likely also possible, and the structure remains stable up to 24.5 GPa before the material loses its long-range crystalline order. Although AgSbTe$_2$ is experimentally known as a semiconducting compound, our electronic structure calculations reveal semimetallic behavior, consistent with previous reports at ambient pressure [13,14,46-50]. Nevertheless, these calculations provide clear and conclusive insights, and we observe an enhancement in the electronic density of states with increasing pressure, which likely contributes to the observed concurrent increase in the $T_c$. Notably, the signature of a dynamic instability between 25 and 30 GPa coincides with the pressure range over which XRD data indicates the onset of loss of the long-range crystalline order. The observation of pressure-induced superconductivity in AgSbTe$_2$, a known thermoelectric material, opens new avenues for exploring superconductivity in related compounds, for expanding the potential applications of this material beyond its already established thermoelectric applications, and for understanding the mechanisms governing superconductivity in chalcogenides. Although the present study establishes the emergence of superconductivity under pressure in Ag$_{0.9}$Sb$_{1.1}$Te$_{2.1}$, further investigations such as high-pressure magnetic measurements and quantitative electron-phonon coupling calculations will be essential to fully clarify the underlying pairing mechanism.

**Credit author statement**

Sudaice Kazibwe: Conceptualization, Investigation, Methodology, Writing– original draft, Writing– review and editing. Bishnu Karki: Investigation, Methodology, Writing– original draft. Wencheng Lu: Investigation, Methodology, Writing– original draft. Zhongxin Liang: Investigation, Methodology, Writing– original draft. Minghong Sui: Formal analysis. Melissa Gooch: Formal analysis, Writing– original draft. Zhifeng Ren: Supervision. Timothy Strobel: Formal analysis, Supervision. Pavan Hosur: Supervision. Ching-Wu Chu: Funding acquisition, Supervision. Liangzi Deng: Funding acquisition, Supervision, Writing–original draft, Writing– review and editing.

**Acknowledgements**

S.K., M.S., M.G., C.W.C, and L.Z.D. in Houston are supported in part by the U.S. Air Force Office of Scientific Research Grants FA9550-15-1-0236 and FA9550-20-1-0068; the T. L. L. Temple Foundation; the John J and Rebecca Moores Endowment; and the State of Texas through the Texas Center for Superconductivity at the University of Houston. L.Z.D. in Houston is supported in part by the Robert A. Welch Foundation. B.K. and P.H. are supported by the state of Texas through the Texas Center for Superconductivity at the University of Houston. W.L. and T.S. acknowledge the GeoSoilEnviroCARS (The University of Chicago, Sector 13), Advanced Photon Source, Argonne National Laboratory for provision of synchrotron radiation facilities at beamline. GeoSoilEnviroCARS is supported by the National Science Foundation – Earth Sciences via SEES: Synchrotron Earth and Environmental Science (EAR –2223273). This research used resources of the Advanced Photon Source, a U.S. Department of Energy (DOE) Office of Science User Facility operated for the DOE Office of Science by Argonne National Laboratory under Contract No. DE-AC02-06CH11357. Use of the GSECARS Raman Lab System was supported by the NSF MRI Proposal (EAR-1531583).

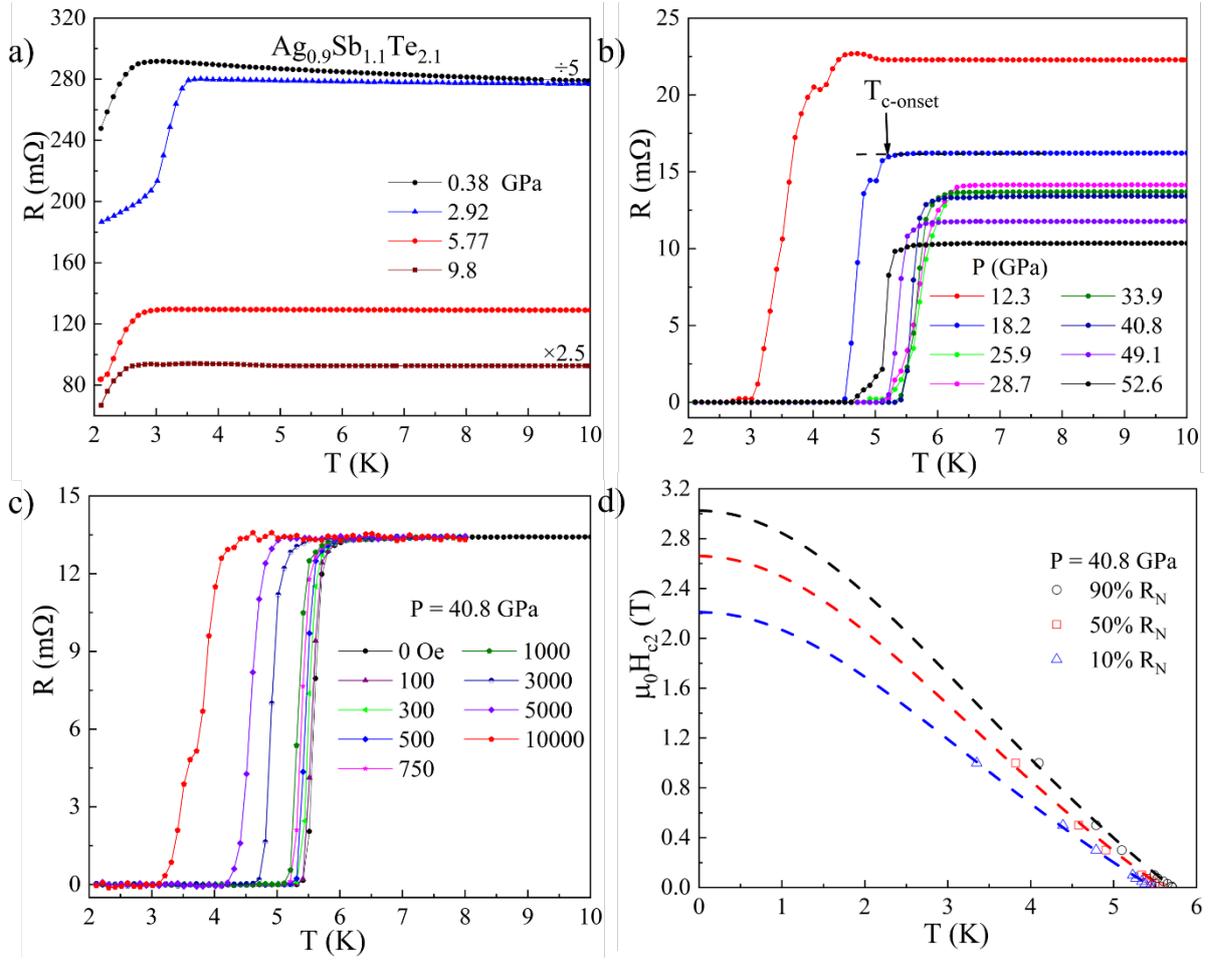

Fig. 1 | Electrical transport measurements of $Ag_{0.9}Sb_{1.1}Te_{2.1}$ under pressure. (a) Resistance–temperature (R–T) plots measured between 2 and 10 K. The resistance at 0.38 GPa was reduced by a factor of 5, and the resistance at 9.8 GPa was increased by a factor of 2.5 for clarity of scale. (b) R–T plots measured from 12.3 GPa up to 52.6 GPa. (c) Temperature-dependent magnetic field-effect measurements at 40.8 GPa, clearly showing a decrease in the $T_c$ with increasing field. (d) Temperature-dependent upper critical field $H_{c2}$ estimated using Ginzburg-Landau fitting at different criteria, corresponding to 90%, 50%, and 10% drops from the normal-state resistance.

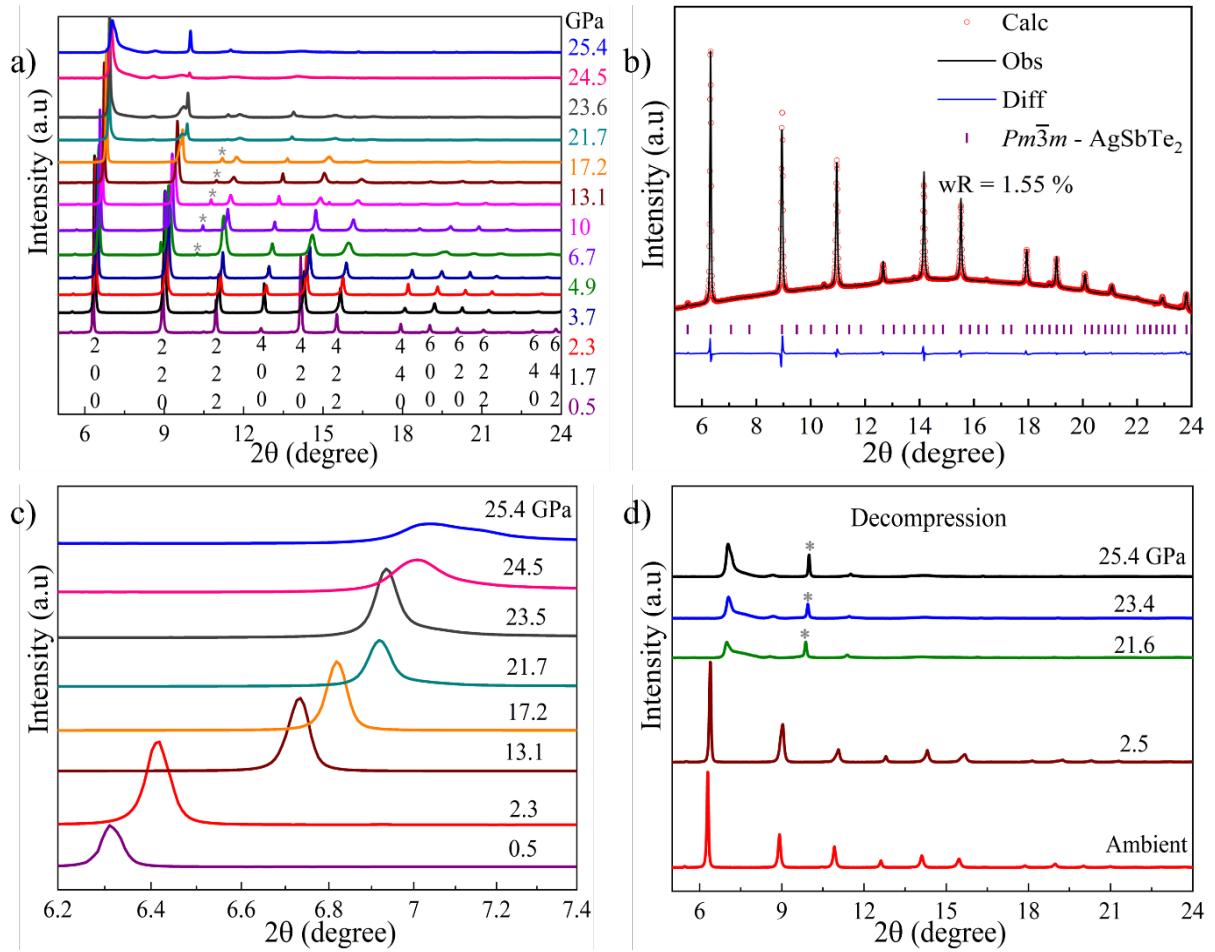

Fig. 2 | Synchrotron XRD patterns of $Ag_{0.9}Sb_{1.1}Te_{2.1}$ measured at room temperature. (a) Evolution of the diffraction patterns upon compression up to 25.4 GPa. The peaks marked with a gray star (*) correspond to the solidification of neon which was used as a PTM. (b) Rietveld refinement of the diffraction pattern of the starting material at 0.5 GPa, where red circles denote the experimental data, and the black and blue curves correspond to the refined fit and difference profile, respectively. (c) Pressure-dependent evolution of the (200) Bragg reflection, showing a systematic shift toward higher 2θ angles with increasing pressure and pronounced peak broadening above 24.5 GPa. (d) XRD patterns collected during decompression, revealing recrystallization of the sample upon recovery to AP. The peaks marked with a gray star (*) result from neon, which is used as a PTM.

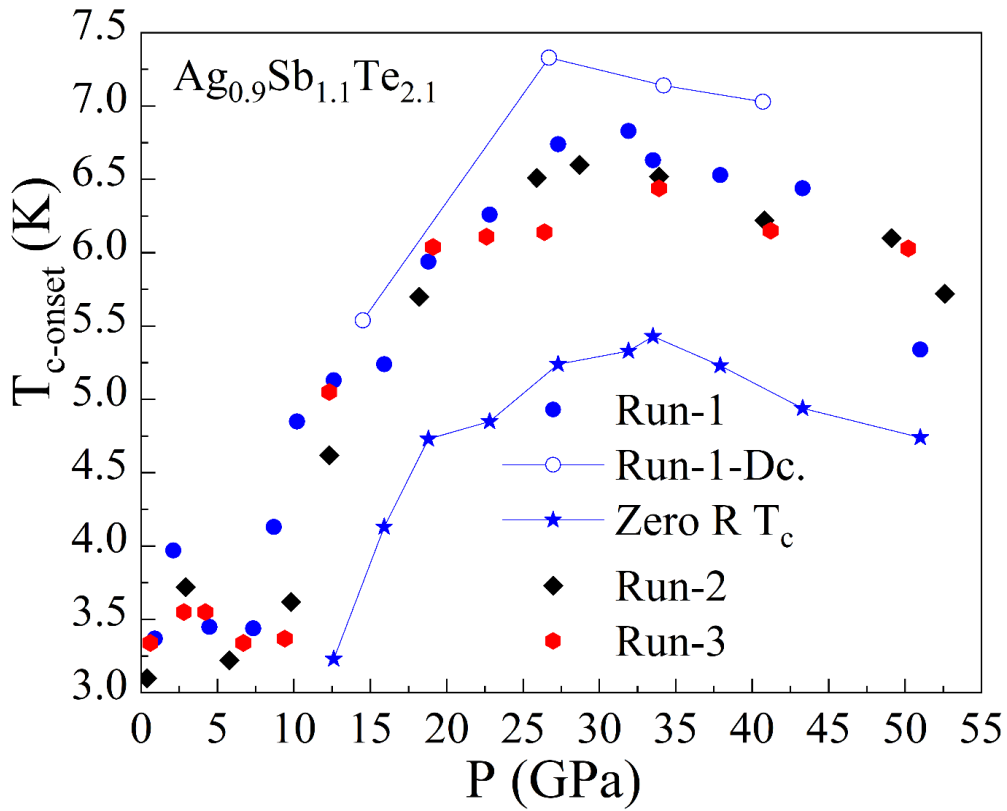

Fig. 3 | Pressure dependence of the superconducting transition temperature up to 55 GPa for $Ag_{0.9}Sb_{1.1}Te_{2.1}$. Solid symbols represent $T_c$ measured during compression and open circles represent $T_c$ measured during decompression (Dc.). Points marked with a star correspond to the zero resistance $T_c$. All runs were performed using c-BN as the PTM.

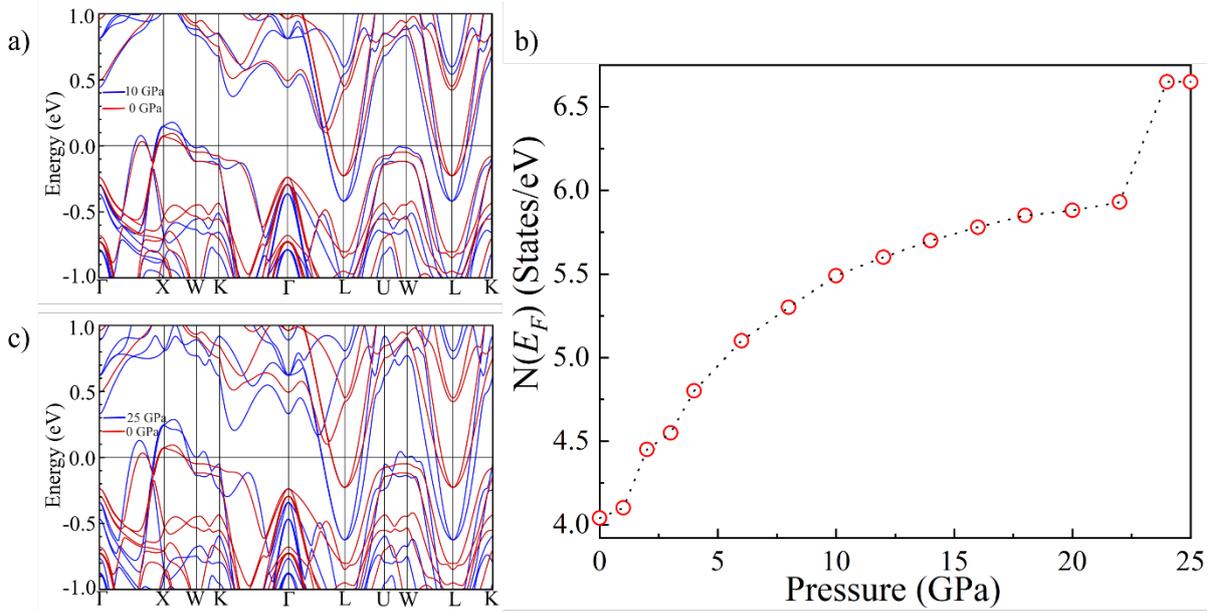

Fig. 4 | Calculated electronic band structure of AgSbTe$_2$ using the ordered $Fd\bar{3}m$ structure with spin-orbit coupling included in the calculations. (a) Band structures calculated at 0 and 10 GPa. (b) Calculated density of states as a function of applied pressure. (c) Band structure at 25 GPa compared with that at ambient pressure.

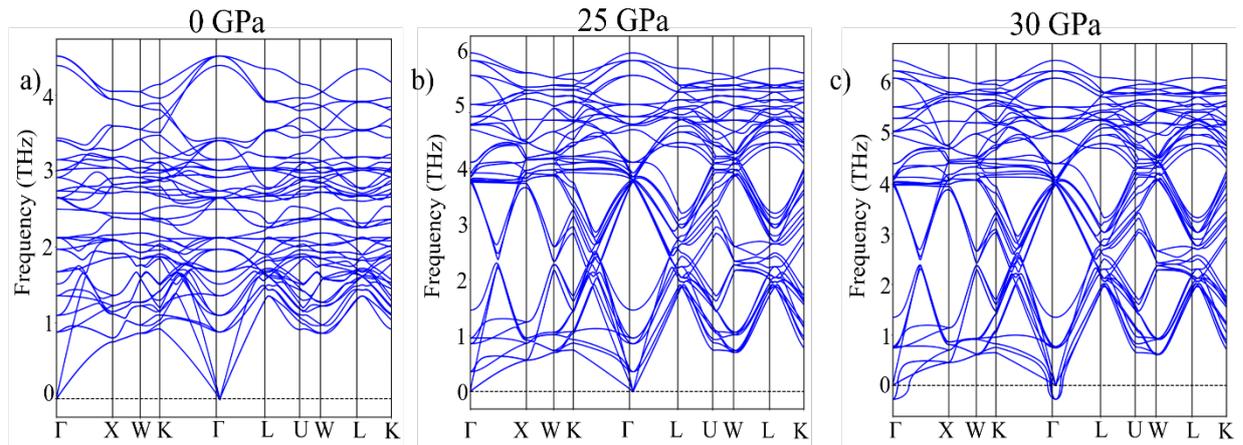

Fig. 5 | Calculated phonon dispersions of the AgSbTe$_2$ system at (a) 0 GPa, (b) 25 GPa, and (c) 30 GPa. The presence of imaginary modes at 30 GPa indicates the structure becomes unstable in the pressure range between 25 and 30 GPa.

# Pressure-induced Superconductivity in AgSbTe$_2$

Sudaice Kazibwe[1], Bishnu Karki[1], Wencheng Lu[2], Zhongxin Liang[1], Minghong Sui[1], Melissa Gooch[1], Zhifeng Ren[1], Pavan Hosur[1], Timothy A. Strobel[2*], Ching-Wu Chu[1*], Liangzi Deng[1*]

1. Department of Physics and Texas Center for Superconductivity, University of Houston, Houston, Texas 77204, USA

2. Earth and Planets Laboratory, Carnegie Institution for Science, Washington, DC 20015, USA

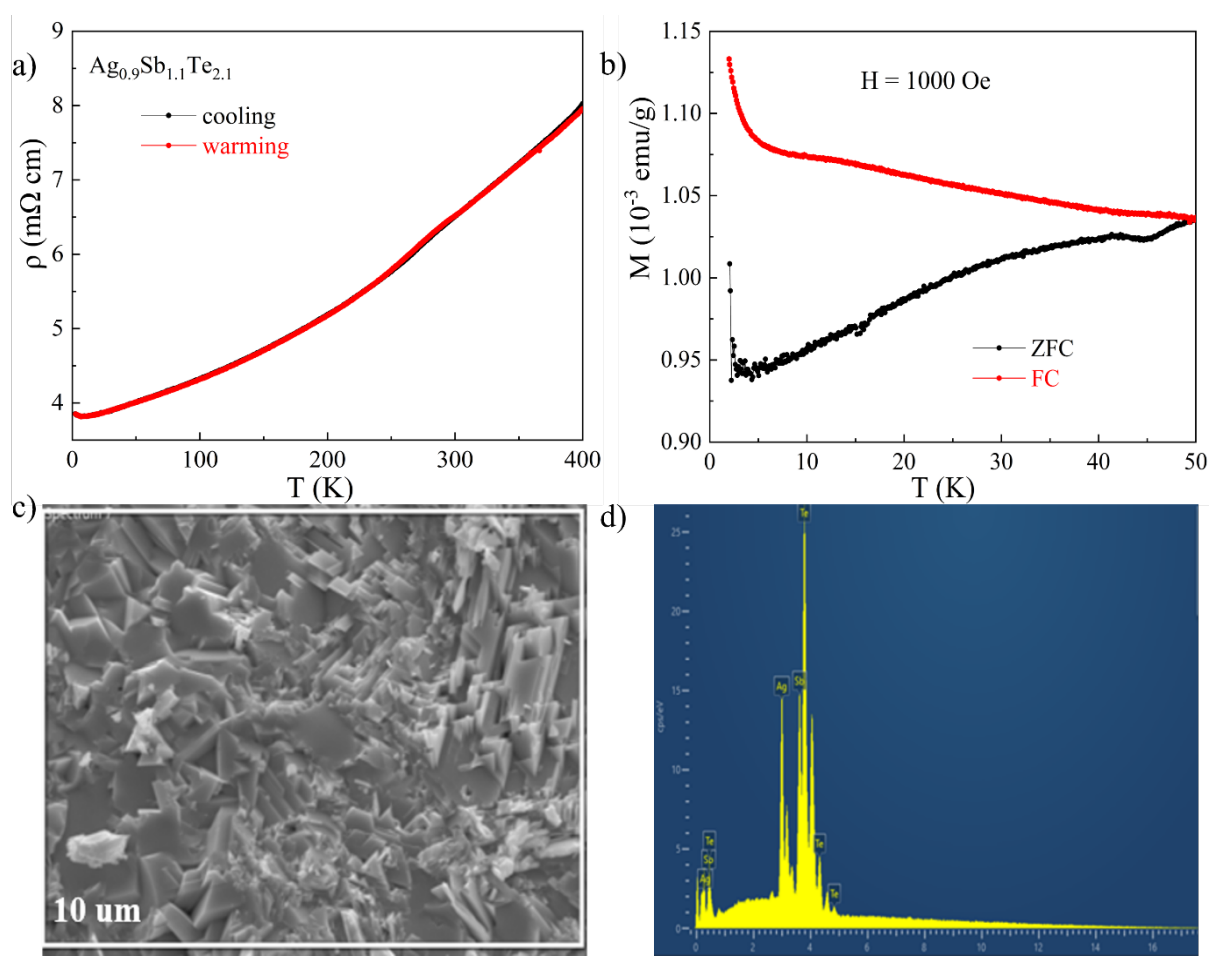

Fig. S1 | Ambient-pressure characterization of Ag$_{0.9}$Sb$_{1.1}$Te$_{2.1}$: (a) resistance–temperature (R–T) measurements from 2 to 400 K show metallic behavior with a low-temperature upturn in resistance; (b) magnetization–temperature (M–T) measurements from 2 to 50 K show no evidence of superconductivity; and (c) an SEM image reveals stacked, angular plate-like particles, from which the EDS spectrum shown in (d) was obtained.

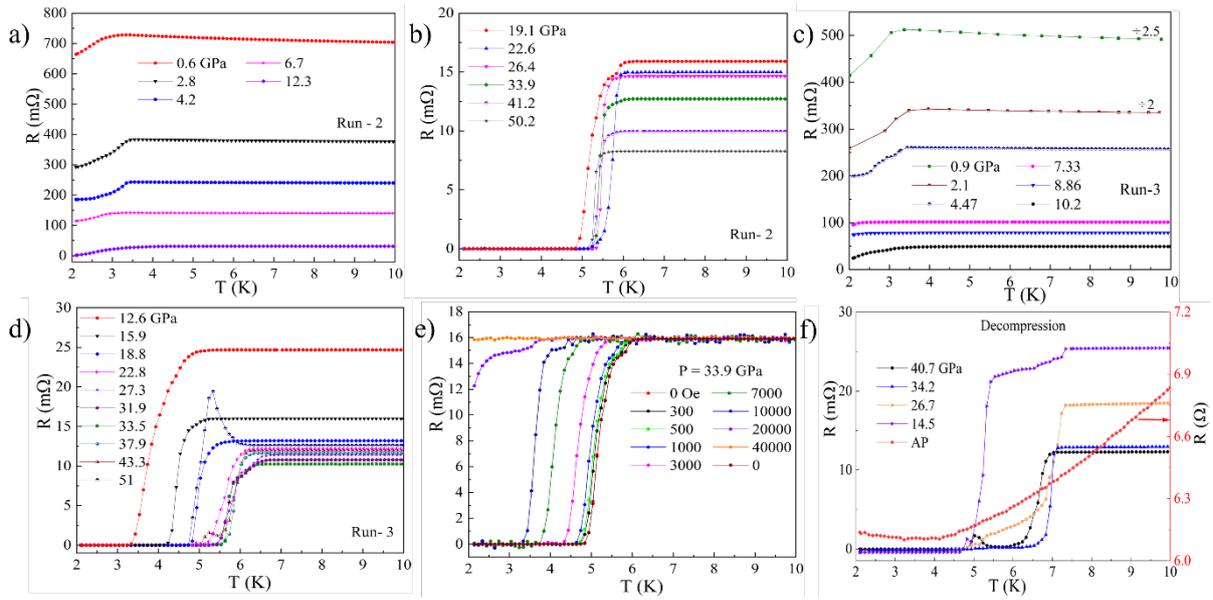

Fig. S2 | Electrical transport measurements under pressure: (a) Resistance–temperature (R–T) measurements up to 12.3 GPa, showing behavior similar to Run 1; measurements were extended up to 50.2 GPa, as shown in (b). (c) and (d) show the R–T plots for Run 3. All runs were performed using c-BN as the PTM. (e) R–T curves measured under different magnetic fields up to 4 T. (f) R–T curves measured during decompression in Run 1, showing suppression of the superconducting phase as pressure is released to ambient conditions.

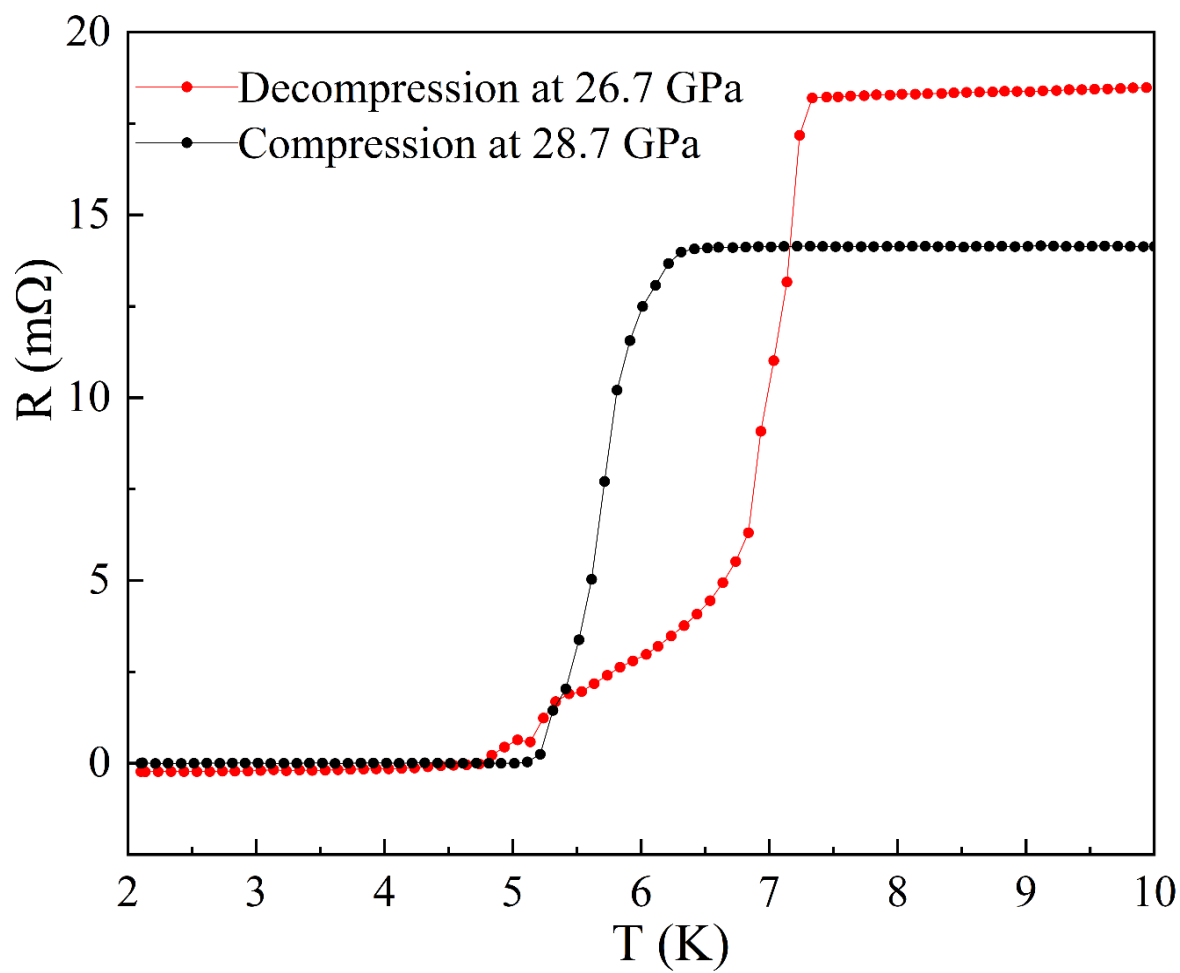

Fig. S3 | Resistance–temperature (R–T) measurements of $Ag_{0.9}Sb_{1.1}Te_{2.1}$ measured during compression at 28.7 GPa and during decompression at 26.7 GPa, clearly demonstrating an enhancement of the superconducting critical temperature upon decompression.

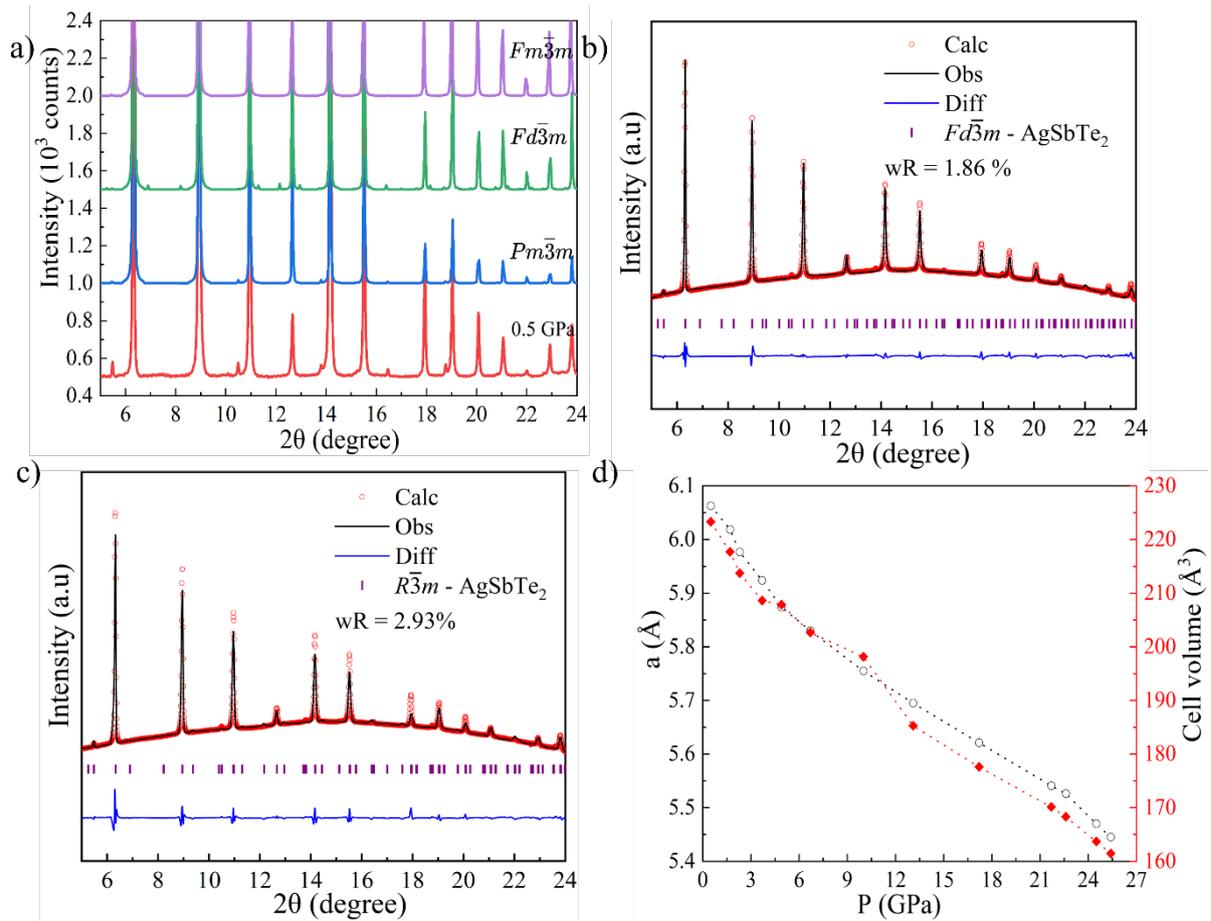

Fig. S4 | XRD patterns of AgSbTe$_2$ at 0.5 GPa. (a) Our sample (red) compared with different cubic space groups. (b) Rietveld refinement using the ordered space group $Fd\bar{3}m$, yielding a weighted profile R-factor of 1.86%. (c) Rietveld refinement using the rhombohedral structure ($R\bar{3}m$). The results show that the sample can be better fitted with both $Pm\bar{3}m$ and $Fd\bar{3}m$, with $Pm\bar{3}m$ (Fig. 2b) giving better results. (d) Pressure dependence of the "a" lattice parameter and cell volume (open black circles and solid red circles, respectively).

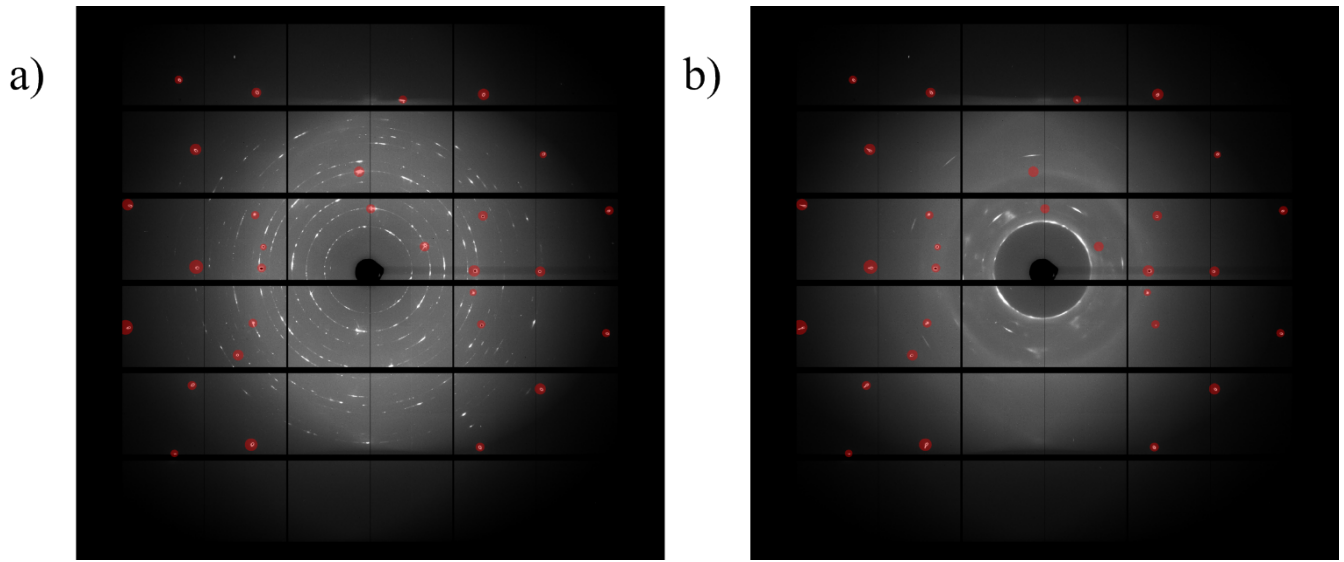

Fig. S5 | 2D diffraction images of $Ag_{0.9}Sb_{1.1}Te_{2.1}$ collected at (a) 0.5 GPa and (b) 25.4 GPa. The red spots indicate masked regions. The data demonstrate a pressure-induced loss of long-range crystalline order when the sample is compressed to 25.4 GPa.

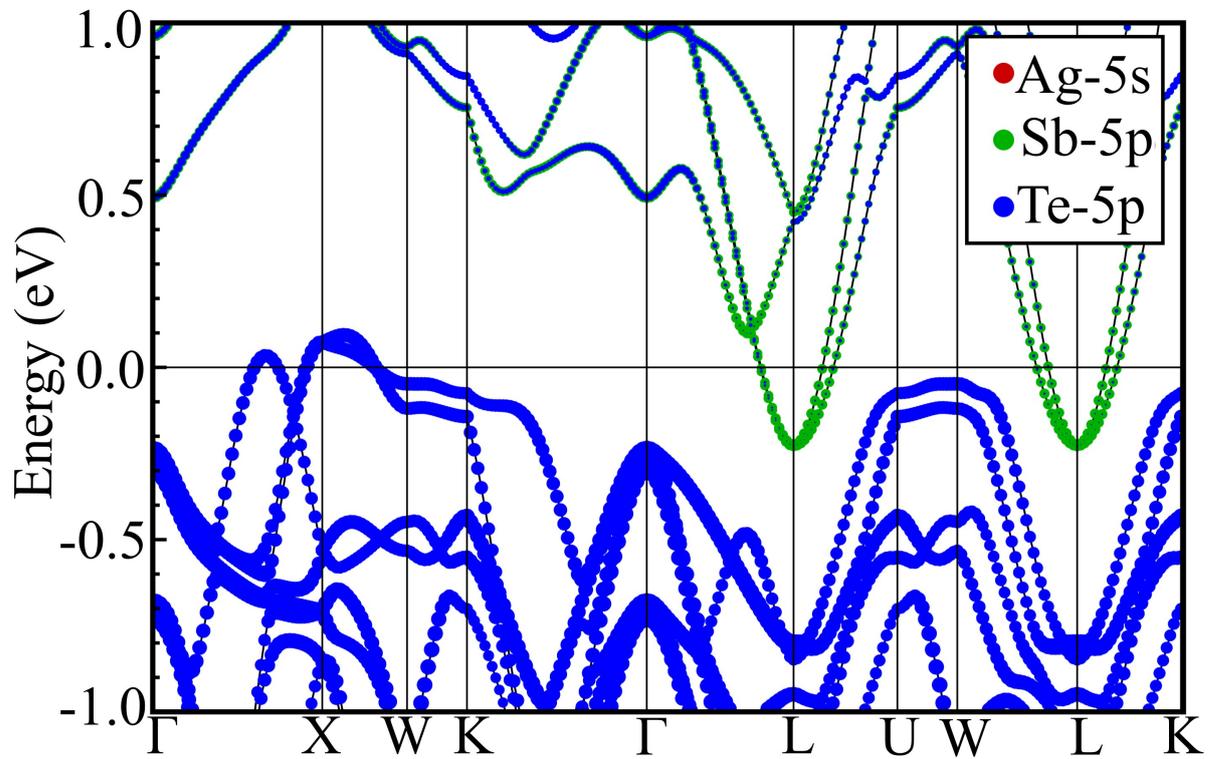

Fig. S6 | Electronic band structure of AgSbTe$_2$ showing that the valence band is predominantly composed of Te-5*p* states, while the conduction band is dominated by Sb-5*p* states. The Sb-5*p* bands cross the Fermi level and overlap with the valence band, revealing the semimetallic nature of the material. The contribution from Ag-5*s* states is minor compared to that of the Sb-5*p* states.

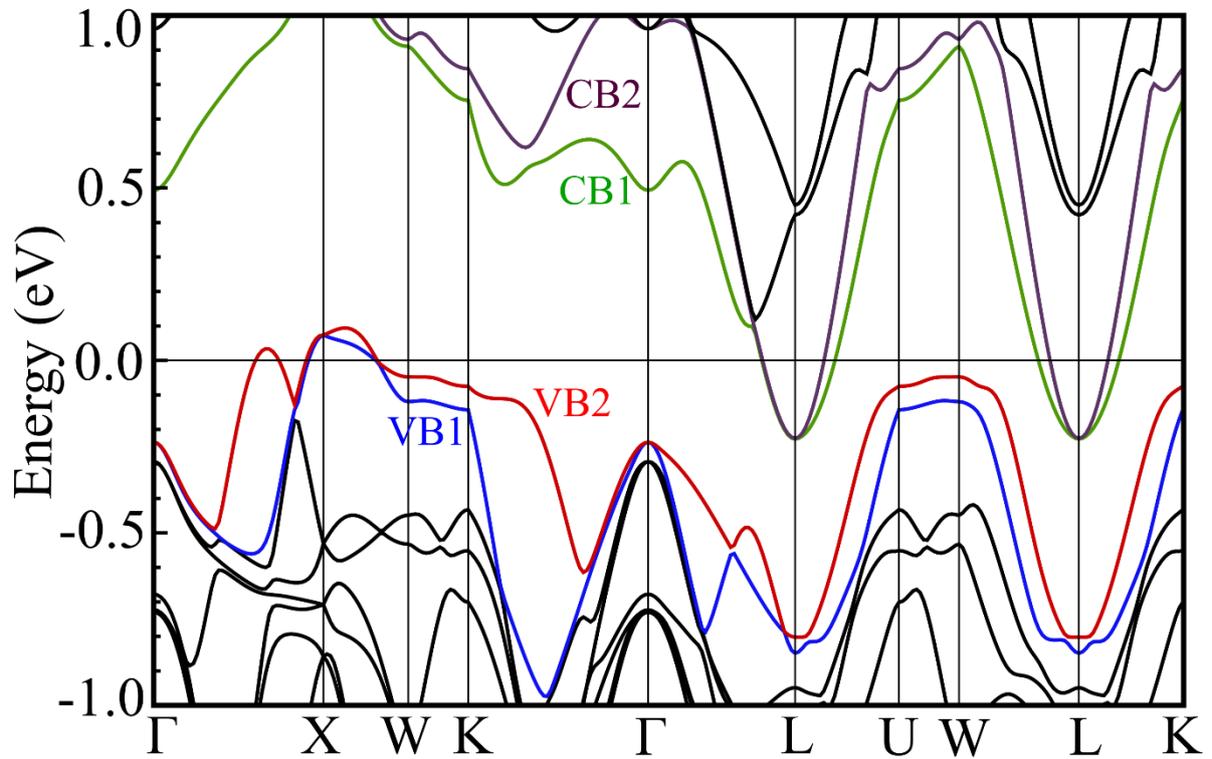

Fig. S7 | Electronic band structure of AgSbTe$_2$ along high-symmetry directions of the Brillouin zone. The valence-band maxima (VB1 and VB2) and conduction-band minima (CB1 and CB2) are highlighted to illustrate the multi-valley nature of the electronic structure. The energy zero is set at the Fermi level.

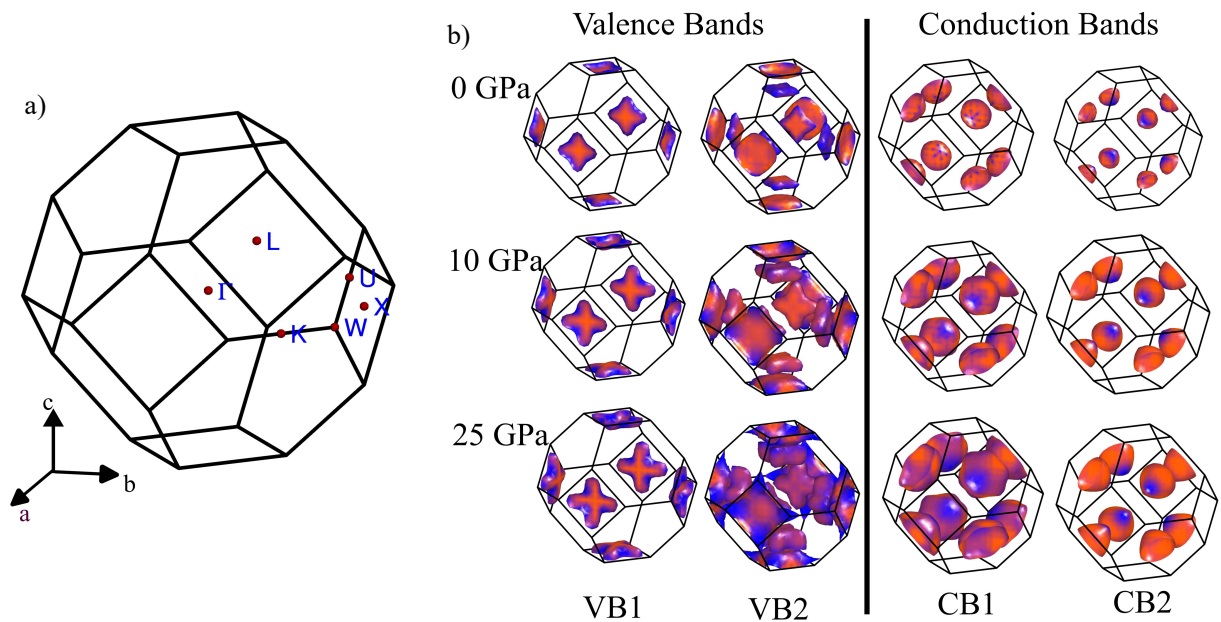

Fig. S8 | Brillouin-zone geometry and pressure evolution of band-edge states in AgSbTe$_2$. (a) First Brillouin zone with high-symmetry k-points and the band-structure path. (b) Constant-energy surfaces of the valence-band maxima (VB1, VB2) and conduction-band minima (CB1, CB2) at 0, 10, and 25 GPa.